\documentclass[aps,nofootinbib,showpacs]{revtex4}
\usepackage{amsfonts}
\def\Lie{\pounds}

\begin{document}

\preprint{gr-qc/0505018}

\title{Quasi-local energy-momentum and energy flux at null infinity}

\author{Xiaoning Wu} \email{wuxn@phy.ncu.edu.tw}
\author{Chiang-Mei Chen} \email{cmchen@phy.ncu.edu.tw}

\affiliation{Department of Physics, National Central University,
Chungli 32054, Taiwan}

\author{James M. Nester} \email{nester@phy.ncu.edu.tw}

\affiliation{Department of Physics and Institute of Astronomy,
National Central University, Chungli 32054, Taiwan}

\date{\today}

\begin{abstract}
The null infinity limit of the gravitational energy-momentum and
energy flux determined by the covariant Hamiltonian quasi-local
expressions is evaluated using the NP spin coefficients. The
reference contribution is considered by three different embedding
approaches. All of them give the expected Bondi energy and energy
flux.
\end{abstract}

\pacs{04.20.-q, 04.20.Cv}

\maketitle

\section{Introduction}
It is well-known that, as a consequence of the equivalence
principle, gravitational energy cannot be localized (see e.g.~
\cite{MTW73} \S 20.4). An alternative idea is quasi-local, namely
quantities associated with a closed two-surface \cite{Pe82}.
During the recent decades there have been numerous intensive
efforts made in the search for a better definition of quasi-local
energy (as well as momentum and angular momentum) for gravitating
systems, with the goal of obtaining quasi-local quantities which
can provide a description of the gravitational field more
elaborate than that given by the total quantities. A very nice
review on the development and applications of quasi-local
quantities in general relativity can be found in \cite{Sz04}. We
are interested in testing certain quasi-local expressions for
energy-momentum and energy flux obtained from the Hamiltonian
boundary term using the covariant Hamiltonian formalism applied to
gravity \cite{CNT95,CN99,CN00,CNT05}.


Some basic criteria are usually presumed for a physically
reasonable definition of quasi-local gravitational energy
\cite{Sz04,LY05}. One of the most important is to consider the
asymptotic behavior of the quasi-local energy when the two surface
approaches null and spatial infinity. The first aim of this work
is to check,  using the NP spin coefficient formalism \cite{PR84},
the null infinity limit of the quasi-local energy-momentum
associated with the boundary of a finite region for a certain
covariant Hamiltonian quasi-local energy-momentum expression.


We are interested in gravitational quantities like energy in the
quasi-local sense, i.e. within a finite region of the space-time.
There is an important issue concerning how much energy flux flows
into and out of the considered region. There continues to be
considerable interest this topic \cite{MF03,Yoon04,BLY97,CJK02}.
In this work we also investigate a natural expression for the
energy flux associated with the aforementioned quasi-local energy
\cite{CNT05}. In order to test whether the definition is suitable,
we again look to the the null infinity limit using the spin
coefficient techniques. In this case, the gravitational energy
flux is expected to be given by the well known Bondi energy loss
formula.


In order to have a reasonable definition of gravitational energy,
the choice of reference plays a essential role. It is well known
that the technique for choosing the reference is an important
unsolved problem for the quasi-local energy issue. The ambiguity
comes from how to embed the reference configuration into the
physical space-time.  Here we consider three different embeddings.


This paper is organized as follows: In the next section, we start
from a basic review on the asymptotic behavior of space-time near
null infinity including a discussion of the expansion of the
Newman-Penrose coefficients which is more complete than the
well-known ones in \cite{PR84,St90,NP62,NT80,GHP73}. Our main
result is contained in three parts: In Section III, we review the
covariant Hamiltonian formalism and the associated boundary term
approach to quasi-local energy momentum and energy flux. We then
re-write the energy-momentum expression in terms of the
Newman-Penrose formalism. In section IV, we find the asymptotic
behavior at null infinity under three different embedding methods.
From the results, we find that the Brown-Lau-York (BLY) embedding
\cite{BLY97} directly gives the standard Bondi mass aspect, while
two other embeddings include  an additional term which, however,
vanishes upon integration. We identify the source of this
difference. In section V, we look at the direct definition of the
energy flux, and consider its null infinity limit to test whether
this definition is reasonable. The detailed calculation is divided
into several subsections. In the first subsection, we calculate
the purely physical part of the energy flux and, in later
subsections, we will consider the three different embedding
methods which have been mentioned. All three types of embedding
give the standard Bondi energy flux.  In section VI we test, using
the spin coefficient formalism in the null infinity limit, the new
Hamiltonian identity based expression for energy flux.  It
directly yields the expected Bondi energy-flux value.  Section VII
includes our concluding discussion.

\section{Asymptotic behavior of space-time near null infinity}
We first review the asymptotic behavior of space-time near null
infinity. Here we are interested in the cases in which the
space-time is asymptotically flat. There have been many intensive
investigations of this subject in the past decades. We will follow
the method initiated by Newman, Penrose, and Tod \cite{NP62,NT80}.
Many additional useful results have been discussed in
\cite{NT80,GHP73,LV83,KTW86,Do92}.

In accord with Penrose's conformal compactification method,
i.e.~the Penrose diagram, we assume that $(\tilde M,\, \tilde
g_{\mu\nu},\, \Omega)$ is the conformal compactified manifold of a
physical space-time $(M,\, g_{\mu\nu})$ via a conformal
transformation $\tilde g_{\mu\nu} = \Omega^2\, g_{\mu\nu}$.
Suppose $S$ is a section of future null infinity ${\cal I}^+$, we
choose {\it Bondi coordinates} near ${\cal I}^+$ in the following
way: $(\theta,\, \phi)$ are spherical coordinates on $S$ and $u$
is the affine parameter generating ${\cal I}^+$ such that $u = 0$
on $S$. If $k^\mu$ is a null vector on ${\cal I}^+$ which is not
tangential, namely $k^\mu \notin T({\cal I}^+)$, then the null
geodesics generated by $k^\mu$ will take the coordinates $(u,\,
\theta,\, \phi)$ into the physical space-time $(M,\, g_{\mu\nu})$.
Moreover, on each null geodesic $\psi(r) = \{ u = {\rm const},\,
\theta = {\rm const},\, \phi = {\rm const} \}$, its affine
parameter $r$, in the sense of the physical metric $g_{\mu\nu}$,
can serve as the forth coordinate (in the neighborhood of ${\cal
I}^+$, we assume the null geodesics are complete). Thus we have
constructed coordinates  ($u,\, r,\, \theta,\, \phi$) for the
physical space-time (they are like Bondi's coordinates except that
$r$ is an affine parameter rather than a luminosity distance, a
small difference asymptotically).

In addition to this coordinate construction, we impose, as has
been introduced in \cite{KTW86,Do92}, certain null frame gauges
choices; stated in terms of the spin coefficients they are
\begin{equation}
\rho - \bar \rho = \mu - \bar \mu = \kappa = \varepsilon = \tau -
\bar \alpha - \beta = \bar \pi - \bar \alpha - \beta = 0.
\label{gauge}
\end{equation}
The fact that $\rho$ and $\mu$ are real functions is insured by
the null frame $m$ and $\bar m$ being always tangent to the two
sphere $r = {\rm const}$ and $u = {\rm const}$. The meaning of the
other gauge choices is that $l$ is a null geodesic and the Lie
transport of both $n$ and $m$ along $l$ are tangent to the sphere.

We use the standard notation for derivatives, $D := l^a \nabla_a$,
$D' := n^a \nabla _a$,  $\delta := m^a \nabla_a$, $\bar \delta :=
\bar m^a \nabla_a$, and the usual complex coordinate on the two
sphere, $\zeta := \cot\frac{\theta}2 \ {\rm e}^{i\phi}$. The null
tetrad can be chosen as
\begin{eqnarray}
l &=& \frac{\partial}{\partial r},
\label{tetl} \\
n &=& \frac{\partial}{\partial u} + U \, \frac{\partial}{\partial
r} + X \, \frac{\partial}{\partial \zeta} + {\bar X} \,
\frac{\partial}{\partial \bar \zeta},
\\
m &=& \xi \, \frac{\partial}{\partial \zeta} + \bar \eta \,
\frac{\partial}{\partial \bar \zeta},
\end{eqnarray}
where $U$ (real) and $X, \xi, \eta$ (complex) are undetermined
functions.

For simplicity, in this paper we will only focus on the vacuum
case. Due to the asymptotic flatness, the behavior of the Weyl
curvature satisfies the ``peeling off theorem'' \cite{PR84,St90},
i.e.
\begin{equation}
\Psi_i \sim O(r^{i-5}), \qquad\qquad i = 0, 1, 2, 3, 4.
\label{asyPsi}
\end{equation}

Near null infinity, we expand all quantities in  Taylor series
with respect to $1/r$. Using the Newman-Penrose (NP) equations,
the asymptotic behavior of the spin-coefficients are
\cite{NT80,Do92}
\begin{eqnarray}
\rho &=& - \frac1{r} - \frac{|\sigma^0|^2}{r^3} + O(r^{-5}),
\label{asyrho} \\
\sigma &=& \frac{\sigma^0}{r^2} + \frac{|\sigma^0|^2 \sigma^0 -
\frac12 \Psi^0_0}{r^4} + O(r^{-5}),
\label{asysigma} \\
\alpha &=& \frac{\alpha^0}{r} + \frac{\bar \sigma^0 \bar \alpha^0
+ {\not\!\partial}_0 \bar \sigma^0}{r^2} + O(r^{-3}),
\label{asyalpha} \\
\beta &=& - \frac{\bar \alpha^0}{r} - \frac{\alpha^0
\sigma^0}{r^2} + O(r^{-3}),
\label{asybeta} \\
\tau &=& \bar \pi = \frac{\bar {\not\!\partial}_0 \sigma^0}{r^2} +
O(r^{-3}), \label{asytau}
\end{eqnarray}
where $\sigma^0(u,\theta,\phi)$ and $\Psi^0_0(u,\theta,\phi)$, the
leading order terms of $\sigma$ (of order $r^{-2}$) and $\Psi_0$
(of order $r^{-5}$), are free functions. Moreover the variable
$\alpha^0$ is an abbreviation for $\alpha^0 :=  \frac1{2\sqrt2}
\zeta$, the spin-weight operator ${\not\!\partial}_0$ is defined
by ${\not\!\partial}_0 := \frac{P}{\sqrt2}
\frac{\partial}{\partial \zeta} + 2 s {\bar \alpha}^0$ acting on a
variable with spin-weight $s$ \cite{GHP73}, and $P := 1 + \zeta
\bar \zeta = \sin^{-2} \frac{\theta}2$. For calculating the energy
flux, we also need the asymptotic expansion for the other NP
coefficients. From the vacuum NP equations, after imposing the
gauge conditions (\ref{gauge}), we have
\begin{equation}
D \gamma = (\tau + \bar \pi) \alpha + (\bar \tau + \pi) \beta +
\tau \pi + \Psi_2.
\end{equation}
Using the results in Eqs. (\ref{asyrho}--\ref{asytau}) and the
asymptotic condition (\ref{asyPsi}), it is straightforward to get
\begin{equation}
\gamma = \frac{\gamma_1}{r^2} + O(r^{-3}) = \frac{- \alpha^0 \bar
{\not\!\partial}_0 \sigma^0 + \bar \alpha^0 {\not\!\partial}_0
\bar \sigma^0 - \frac12 \Psi_2^0}{r^2} +
O(r^{-3}).\label{asygamma}
\end{equation}

In order to  get the asymptotic expansion of the remaining NP
coefficients we need to have some more control on the null tetrad.
From the commutation relations, one can easily derive the null
tetrad control equations
\begin{eqnarray}
D U &=& - (\gamma + \bar \gamma),
\label{DU} \\
D X &=& (\bar \tau + \pi) \xi + (\tau + \bar \pi) \eta,
\label{Dx} \\
D \xi &=& \bar \rho \xi + \sigma \eta,
\label{Dxi} \\
D \eta &=& \bar \sigma \xi + \rho \eta,
\label{Deta}
\end{eqnarray}
which lead to the following asymptotic behavior of the null tetrad
\begin{eqnarray}
U &=& - \frac12 + \frac{\gamma_1 + \bar \gamma_1}{r} + O(r^{-2}),
\label{asyU} \\
X &=& - \frac{\xi^0 {\not\!\partial}_0 \bar \sigma^0 }{r^2} +
O(r^{-3}),
\label{asyX} \\
\xi &=& \frac{\xi^0}{r} + \frac{|\sigma^0|^2 \xi^0}{r^3} +
O(r^{-5}),
\label{asypxi} \\
\eta &=& - \frac{\bar \sigma^0 \xi^0}{r^2} + O(r^{-4}),
\label{asyeta}
\end{eqnarray}
where the numerical factor $-\frac12$ in (\ref{asyU}) and the
value of $\xi^0$ ($\xi^0 := \frac{P}{\sqrt2}$) are specified by
the result from Minkowski space-time. The asymptotic behavior of
$\mu$ and $\lambda$ can now be retrieved from the following NP
equations
\begin{eqnarray}
D \mu - \delta \pi &=& (\bar \rho \mu + \sigma \lambda) + |\pi|^2
- (\bar \alpha - \beta) \pi + \Psi_2,
\\
D \lambda - \bar \delta \pi &=& (\rho \lambda + \bar \sigma \mu) +
\pi^2 + (\alpha - \bar \beta) \pi,
\\
\delta \tau - D' \sigma &=& (\mu \sigma + \bar \lambda \rho) +
\tau^2 - (\bar \alpha - \beta) \tau - (3 \gamma - \bar \gamma)
\sigma.
\end{eqnarray}
The last equation is needed to determine the value of the leading
order term of $\lambda$. Using the obtained results
(\ref{asyU}--\ref{asyeta}) and the property that the spin weight
of $\pi$ is $-1$, it is straightforward to get
\begin{eqnarray}
\mu &=& - \frac1{2r} - \frac{\Psi^0_2 + \sigma^0 \dot {\bar
\sigma}{}^0 + {\not\!\partial}_0^2 {\bar \sigma}^0}{r^2} +
O(r^{-3}),
\label{asymu} \\
\lambda &=& \frac{\dot {\bar\sigma}{}^0}{r} + \frac{\frac12 \bar
\sigma^0 - \bar {\not\!\partial}_0 {\not\!\partial}_0 \bar
\sigma^0}{r^2} + O(r^{-3}), \label{asylambda}
\end{eqnarray}
where the dot means derivative with respect to $u$ and the
numerical factor $-\frac12$ in the leading term of $\mu$ is
specified by the Minkowski space-time result.  (It will turn out
that $\mu$ and $\lambda$ will play the key roles in our results.)

Finally the expansion for the coefficient $\nu$ can be obtained
from the NP equation
\begin{equation}
D \nu - D' \pi = (\pi + \bar \tau) \mu + (\bar \pi + \tau) \lambda
+ (\gamma - \bar \gamma) \pi + \Psi_3,
\end{equation}
which gives
\begin{equation}
\nu = - \frac{{\not\!\partial}_0 \dot {\bar \sigma}^0 +
\Psi^0_3}{r} + O(r^{-2}). \label{asynu}
\end{equation}
However, the leading term of the NP equation
\begin{equation}
\delta \lambda - \bar \delta \mu = (\alpha + \bar \beta) \mu +
(\bar\alpha - 3 \beta) \lambda - \Psi_3,
\end{equation}
shows that $\Psi^0_3 = - {\not\!\partial}_0 \dot {\bar \sigma}^0$
(the spin weight of $\bar \sigma^0$ is $-2$). Therefore the
leading order of $\nu$ indeed is $r^{-2}$ which does not make a
contribution in the later calculation.

Moreover, for the later calculation, we still need the asymptotic
properties of the induced volume element (2-dimensional) on the
considered sphere $S^2$. The induced metric ${}^{(2)}ds^2$ on the
sphere $S=\{u={\rm const}, t={\rm const}\}$ asymptotically should
be
\begin{equation}
{}^{(2)}ds^2 = r^2 (d\theta^2 + \sin^2\theta \, d\phi^2) + O(r).
\label{g2}
\end{equation}
The induced volume element on $S^2$ can be obtained from the
volume element $\epsilon :=  i \, {\bf l} \wedge {\bf n} \wedge
{\bf m} \wedge {\bf \bar m}$, in our case $\epsilon = \sqrt{-g} \,
du \wedge dr \wedge d\theta \wedge d\phi$, by
\begin{eqnarray}
{}^2\epsilon_{cd} &=& \epsilon_{abcd} \left(
\frac{\partial}{\partial u} \right)^a
\left(\frac{\partial}{\partial r} \right)^b
\nonumber \\
&=& i \, {\bf m} \wedge {\bf \bar m} + O(r) \nonumber \\
&=& r^2 \left( 1 - \frac{|\sigma^0|^2}{r^2} \right) \sin\theta \,
d\theta \wedge d\phi + O(r^{-2}). \label{vol2}
\end{eqnarray}
In the derivation we have used the relation
$\frac{\partial}{\partial \zeta} = - \sin^2\frac{\theta}2 \, {\rm
e}^{-i\phi} (\frac{\partial}{\partial \theta} +
\frac{i}{\sin\theta} \frac{\partial}{\partial \phi})$.

\section{Quasi-local energy-momentum and its null infinity limit}
There have been many proposals regarding quasi-local quantities.
It should be noted that there is as yet no consensus regarding
what approach should be used or even what are the proper criteria
\cite{Sz04}. However it has been argued that the Hamiltonian
approach, which has been used by many researchers and which we
adopt here, has certain merits (see
e.g.~\cite{CNT95,CN99,CN00,BLY97,BLY98,CNC99,Ne04,CJK02}). For a
general region $\Sigma$ (finite or infinite) the Hamiltonian,
\begin{equation}
H(N,\Sigma) = \int_\Sigma N^a {\cal H}_a + \oint_S {\cal B}(N),
\label{ham}
\end{equation}
which displaces the region along a vector field $N$, includes not
only an integral of a density over the 3-dimensional region but
also an integral over its closed 2-surface boundary
$S=\partial\Sigma$. For Einstein's general relativity (as well as
any other geometric gravity theory) the Hamiltonian densities
${\cal H}_a$ are proportional to certain field equations---the
initial value constraints---and so the {\em value} of the
Hamiltonian, $E(N,S)$, is determined purely by the integral of the
boundary term.  For appropriate choices of the displacement $N$ on
the boundary, this Hamiltonian boundary term, for any gravitating
system, determines the quasi-local values: in particular from a
suitable time-like translation, the quasi-local energy and from a
suitable spacelike translation, the quasi-local linear momentum.
The approach is quite general; it can incorporate in particular
not only all the Noether charge expressions but also all the
traditional pseudotensor expressions (and thereby it rehabilitates
this often discredited approach) while taming the notorious
ambiguities: the choice of boundary expression is linked, via the
boundary term in the variation of the Hamiltonian, with the choice
of boundary condition, while the reference frame ambiguity can be
associated with a choice of boundary reference values, which
determine the choice of vacuum or ground state for the system. In
this way the traditional ambiguities can be given a clear physical
and geometric significance \cite{CNC99,CN00,Ne04}. Within the
covariant Hamiltonian formalism certain {\em covariant symplectic}
expressions for the conserved gravitational quantities have been
proposed \cite{CNT95,CN99,CN00}. When we look at GR from this
perspective one of these expressions stands out as being suitable
for most applications, (among other virtues it has an associated
positive energy proof \cite{Ne89}). Here we consider only this
particular expression, specifically, in geometric units
\begin{equation}
E(N,S) = \frac1{16\pi} \oint_S \left( \Delta \omega^{ab} \wedge
i_N \eta_{ab} + {\stackrel\circ\nabla}{}^b {\stackrel\circ N}{}^a
\Delta \eta_{ab} \right), \label{Energy}
\end{equation}
where $\Delta\omega^{ab}:=\omega^{ab}-\stackrel\circ\omega{}^{ab}$
is the difference between the orthonormal frame connection
one-forms (i.e., the Ricci rotation one-forms) and their reference
values, $\stackrel\circ\nabla$ and ${\stackrel\circ N}{}^a$ are
the connection and the displacement vector in the reference
space-time, $\eta_{ab} := (1/2)\epsilon_{abcd} \, \vartheta^c
\wedge \vartheta^d$, $\Delta \eta_{ab} := (1/2)\epsilon_{abcd} (
\vartheta^c \wedge \vartheta^d - {\stackrel\circ\vartheta}{}^c
\wedge {\stackrel\circ\vartheta}{}^d )$, with $\vartheta^a$ and $
{\stackrel\circ\vartheta}{}^a$ being, respectively, the dynamic
and reference orthonormal co-frames.

The physical and geometric significance of this particular choice
of Hamiltonian boundary term expression is revealed by the
resultant boundary term in the variation of the Hamiltonian:
\begin{equation}
\delta H(N,\Sigma) = \int_\Sigma (\hbox{\rm field equation terms})
- {1\over16\pi} \oint_S i_N (\Delta \omega^{ab} \wedge
\delta\eta_{ab}). \label{deltaH}
\end{equation}
This indicates that we should hold fixed on the boundary $S$ the
pullback of $\eta_{ab}$, i.e., certain projected components of the
coframe---thus, effectively, certain projected components of the
metric (arguably the most natural choice).

In the works already cited this quasilocal expression has been
tested in various ways.  Here we are concerned with the
requirement that the value of the quasi-local energy-momentum and
the energy-flux have the correct limit at null infinity. To this
end we take $S=\partial\Sigma$, the closed boundary of the three
dimensional space-like region $\Sigma$, to be a two sphere which
approaches in the limit  null infinity.  The Hamiltonian formalism
with a boundary approaching null infinity has been considered from
several perspectives, see e.g. \cite{HN96,BLY97}.  A nice detailed
discussion of the topic, addressing all of the important issues
has been given recently \cite{CJK02}.

By choosing the vector $N$, one can derive the 10 conserved
quasi-local quantities for gravity based on the Poincar\'e
symmetry. In particular the covariant quasi-local energy and
momentum associated with the time and space translation asymptotic
symmetries are
\begin{equation}
p_\nu = \frac1{16\pi} \int_S \left( \Delta \omega^{ab} \wedge
i_{N_\nu} \eta_{ab} + {\stackrel\circ\nabla}{}^b {\stackrel\circ
N}{}^a_\nu \Delta \eta_{ab} \right). \label{pnu}
\end{equation}
Here the value of $\nu$ labels what quantities are evaluated, $0$
for energy and $1,2,3$ for the three components of momentum. In
the asymptotically flat case, $N_\nu$ should be the translation
part of the asymptotic Killing vectors. The translation part of
the Bondi-Metzner-Sachs(BMS) group is well-defined; its expansion,
in the leading order, is of the form $N_\nu = N_\nu^{(0)} +
O(r^{-1})$, with $\nu = 0$, $k=1,2,3$:
\begin{eqnarray} \label{expN}
N_0^{(0)} &=& \frac{\partial}{\partial u} = f_0 \left( n + \frac12
\, l \right) + O(r^{-1}),
\nonumber\\
N_k^{(0)} &=& f_k \left( \frac{\partial}{\partial u} -
\frac{\partial}{\partial r} \right) = f_k \left( n - \frac12 \, l
\right) + O(r^{-1}),
\end{eqnarray}
where $f_\nu = (1, -\sin\theta \cos\phi, -\sin\theta \sin\phi,
-\cos\theta)$.

The energy-momentum expression (\ref{pnu}) includes two parts
which will be considered separately: the purely physical part and
the part including the reference, i.e.
\begin{equation}
p_\nu = p_\nu^{\rm phy} + p_\nu^{\rm ref},
\end{equation}
where
\begin{equation}
p_\nu^{\rm phy} := \frac1{16\pi} \int_S \left( \omega^{ab} \wedge
i_{N_\nu} \eta_{ab} \right), \qquad p_\nu^{\rm ref} :=
\frac1{16\pi} \int_S \left( - {\stackrel\circ\omega}{}^{ab} \wedge
i_{N_\nu} \eta_{ab} + {\stackrel\circ\nabla}{}^a {\stackrel\circ
N}{}^b_\nu \Delta \eta_{ab} \right).
\end{equation}

We first evaluate the physical part and leave the reference part
and the final results to the next section. It is important to keep
in mind that the integral is evaluated on a two sphere with
constant $u$ and $r$, therefore the only contributing term is
${\bf m} \wedge {\bf \bar m}$. The gauge (\ref{gauge}) guarantees
that the vector $m$ is always tangent to the 2-sphere. Hereafter,
we will only present the coefficient  of the 2-form ${\bf m}
\wedge {\bf \bar m}$, denoting this as ``$\cong$''.

From the expansion of the vector $N$, (\ref{expN}), we realize
that the the significant contribution for $N = f_\nu \left( n \pm
\frac12 \, l \right)$ is
\begin{eqnarray}
\omega^{ab} \wedge i_{N_\nu} \eta_{ab} &=& f_\nu \left\{ - i 2
(\gamma - \bar \gamma) \, {\bf l} \wedge {\bf n} \pm i (\alpha -
\bar \beta + \bar \tau \pm 2 \nu) \, {\bf l} \wedge {\bf m} \mp i
(\bar \alpha - \beta + \tau \pm 2 \bar \nu) \, {\bf l} \wedge {\bf
\bar m} \right.
\nonumber\\
&& \left. - i 2 (\alpha - \bar \beta - \pi) \, {\bf n} \wedge {\bf
m} + i 2 (\bar \alpha - \beta - \bar \pi) \, {\bf n} \wedge {\bf
\bar m} + i \left[ 2(\mu + \bar \mu) \pm (\rho + \bar \rho)
\right] \, {\bf m} \wedge {\bf \bar m} \right\}. \label{Freud}
\end{eqnarray}
The imaginary unit $i$ comes from the volume element, i.e.
$\epsilon = i \, {\bf l} \wedge {\bf n} \wedge {\bf m} \wedge {\bf
\bar m}$ \cite{KSHM80}. However, if we only focus on the value on
the two sphere boundary, the result is
\begin{equation}
\omega^{ab} \wedge i_{N_\nu} \eta_{ab} \cong i f_\nu \left[ 2(\mu
+ \bar \mu) \pm (\rho + \bar \rho)  \right] \, {\bf m} \wedge {\bf
\bar m}. \label{EMphy}
\end{equation}

\section{Gravitational energy-momentum in different embeddings}
Reference configurations play a crucial role in the expressions
for gravitational energy  and its energy flux (indeed for all the
quasi-local quantities). There are two essential related issues:
(i) a ``suitable'' reference configuration choice and (ii) a
proper embedding into the physical space-time.

There are two terms in the reference part. The NP formulation for
the first one, ${\stackrel\circ\omega}{}^{ab} \wedge i_N
\eta_{ab}$ can be easily read out from (\ref{Freud}) by replacing
all NP coefficients and the frames within the connection with
their reference values.

For an asymptotically flat space-time the choice of reference
configuration is more or less unambiguous --- the Minkowski
space-time. In Eddington-Finkelstein coordinates ($u,\, r,\,
\theta,\, \phi$), which are related to the standard coordinates by
$u = t - r$, the first fundamental form of the Minkowski
space-time is
\begin{equation}
ds^2 = - du^2 - 2 du dr + r^2 (d\theta^2 + \sin^2\theta d\phi^2).
\end{equation}
It is easy to verify that the coordinates are just the Bondi
coordinates.

However, the issue of embedding the two surface $S$ of physical
space-time into the Minkowski space-time is not fully transparent
yet. Various proposals have been made in earlier works. In this
subsection, we consider several types of embedding used for
calculating the reference part of the energy; later we will also
use these embeddings for energy flux. We will see that the
embeddings we consider actually give the same result for the
gravitational energy flux at future null infinity, but not quite
the same expression for energy itself.

\subsection{Holonomic embedding}
The simplest embedding technique is to identify the
Bondi-coordinates with the Minkowski coordinates, i.e. the
embedded surface $S_0$ is just the standard coordinate round
sphere. In this approach, the embedded null frame is
\begin{eqnarray}
{\stackrel\circ l} &=& \frac{\partial}{\partial r}, \\
{\stackrel\circ n} &=& \frac{\partial}{\partial u} - \frac12
\frac{\partial}{\partial r}, \\
{\stackrel\circ m} &=& \frac{{\rm e}^{-i\phi}}{\sqrt2 r} \left(
\frac{\partial}{\partial \theta} + \frac{i}{\sin\theta}
\frac{\partial}{\partial \phi} \right) = \frac{P}{\sqrt2 r}
\frac{\partial}{\partial \zeta}.
\end{eqnarray}
Therefore the non-vanishing N-P reference coefficients are
\begin{equation}
{\stackrel\circ\rho} = - \frac1{r}, \qquad {\stackrel\circ\mu} = -
\frac1{2r}, \qquad {\stackrel\circ\alpha} = -
{\stackrel\circ{\bar\beta}} =  \frac{\zeta} {2\sqrt2 \, r}.
\end{equation}
The embedding of the time direction ${\stackrel\circ N}$ is
${\stackrel\circ N}_\nu = f_\nu \left( {\stackrel\circ n} + \frac12
\, {\stackrel\circ l} \right)$ which is a Killing vector of the
Minkowski space-time.   Properly we should calculate
${\stackrel\circ\omega}{}^{ab}=
\stackrel\circ\Gamma\!\!{}^{ab}{}_c\stackrel\circ\vartheta\!\!{}^c$.
However since the difference of the coframes  $\vartheta^c$ and
$\stackrel\circ\vartheta\!{}^c$ is $o(r^{-1})$ we can make an
approximation and take for the reference part
\begin{equation}
\stackrel\circ\omega{}^{ab} \wedge i_N \eta_{ab} \cong i f_\nu
\left[ 2( \stackrel\circ\mu + \stackrel\circ{\bar\mu} ) \pm (
\stackrel\circ\rho + \stackrel\circ{\bar\rho} ) \right] \, {\bf m}
\wedge {\bf \bar m}. \label{EMref1a}
\end{equation}
Moreover, it is straightforward to check, for $\stackrel\circ
N_\nu = f_\nu \left( \stackrel\circ n \pm \frac12 \stackrel\circ l
\right) + O(r^{-1})$, that
\begin{eqnarray}
{\stackrel\circ\nabla}{}^a {\stackrel\circ N}{}^b_\nu \Delta
\eta_{ab} &=& ( {\stackrel\circ\partial}{}^a f_\nu ) (
\stackrel\circ n \pm \frac12 \stackrel\circ l )^b \Delta \eta_{ab}
+ f_\nu {\stackrel\circ\nabla}{}^a ( \stackrel\circ n \pm \frac12
\stackrel\circ l )^b  \Delta \eta_{ab} + \stackrel\circ\nabla
O(r^{-1}) \Delta \eta
\nonumber\\
&=& ( {\stackrel\circ\partial}{}^a f_\nu ) ( \stackrel\circ n \pm
\frac12 \stackrel\circ l )^b \Delta \eta_{ab} - i f_\nu \left[ (
\stackrel\circ\gamma + \stackrel\circ{\bar\gamma} ) \Delta ( {\bf
m} \wedge {\bf \bar m} ) + \stackrel\circ\nu \Delta ( {\bf l}
\wedge {\bf m} ) - \stackrel\circ{\bar\nu} \Delta ( {\bf l} \wedge
{\bf \bar m} ) \right] + O(r^{-1})
\nonumber\\
&\cong& O(r^{-1}). \label{EMref1b}
\end{eqnarray}

Finally, from the results (\ref{EMphy}, \ref{EMref1a},
\ref{EMref1b}), the energy-momentum in the holonomic embedding is
\begin{eqnarray}
\lim_{\cal I^+}p_\nu &=& \lim_{\cal I^+}\frac1{16\pi} \int_S f_\nu
\left[ 2( \mu + \bar \mu - \stackrel\circ\mu -
\stackrel\circ{\bar\mu} ) \pm ( \rho + \bar \rho -
\stackrel\circ\rho - \stackrel\circ{\bar\rho} ) \right] \, i \ {\bf
m} \wedge {\bf \bar m}
\nonumber\\
&=& - \frac1{4 \pi} \int_S {\rm Re} \left( \Psi^0_2 + \sigma^0
\dot {\bar \sigma}{}^0 + {\not\!\partial}_0^2 {\bar \sigma}^0
\right) f_\nu \, d\Omega^2,
\end{eqnarray}
where $d\Omega^2 = \sin\theta \, d\theta d\phi$. The integrand
differs slightly from the usual formula for the Bondi
energy-momentum; however the integral of the extra term vanishes
at least for the energy---which is our real interest here---simply
because ${\not\!\partial}{\bar \sigma}^0$ is of spin weight -1.
Note that entirely analogous terms show up in equivalent
calculations done directly in the Bondi-Sachs metric
\cite{HN96,CNT05}.

\subsection{\'O Murchadha-Szabados-Tod embedding}
In Ref. \cite{OST04}, \'O Murchadha, Szabados and Tod (OST)
introduced another kind of embedding method. Let us consider the
space-like region $\Sigma_0$ in the physical space-time, $\partial
\Sigma_0 = S$. We suppose that $S$ is a topological two sphere and
that isothermal coordinates globally exist on $S$. In these
coordinates, the induced metric ${}^{(2)}ds^2$ on $S$ is
\begin{equation}
{}^{(2)}ds^2 = \omega^2(\theta',\phi') (d\theta'^2 + \sin^2\theta'
d\phi'^2),
\end{equation}
where $\omega(\theta',\phi')$ is a positive function on $S$. This
two surface is embedded isometrically into the physical space-time
(in Bondi coordinates) with
\begin{equation}
u = {\rm const}, \qquad \theta = \theta', \qquad \phi = \phi',
\qquad r = \omega(\theta,\phi).
\end{equation}
Based on this embedding, we define a coordinate transformation in
Minkowski space-time,
\begin{equation}
u \to U, \qquad r \to R + \omega, \qquad \theta \to \theta, \qquad
\phi \to \phi.
\end{equation}
Then the Minkowski metric becomes
\begin{equation}
{\stackrel\circ g} = \left( \begin{array}{cccc}
-1 & -1 & - \partial_\theta \omega & - \partial_\phi \omega \\
-1 & 0 & 0 & 0\\
- \partial_\theta \omega & 0 & (R + \omega)^2 & 0 \\
- \partial_\phi \omega & 0 & 0 & (R+\omega)^2 \sin^2\theta
\end{array} \right).
\end{equation}

The NP reference tetrad is chosen to be
\begin{eqnarray}
\stackrel\circ l &=& \frac{\partial}{\partial R},
\nonumber \\
\stackrel\circ n &=& \frac{\partial}{\partial U} - \frac12 \left(1
+ \frac{|\delta_0 \omega|^2}{(R + \omega)^2} \right)
\frac{\partial}{\partial R} + \frac{\partial_\theta \omega}{(R +
\omega)^2} \frac{\partial}{\partial \theta} + \frac{\partial_\phi
\omega}{(R + \omega)^2 \sin^2\theta} \frac{\partial}{\partial
\phi},
\nonumber \\
\stackrel\circ m &=& \frac{{\rm e}^{-i\phi}}{\sqrt2 (R + \omega)}
\left( \frac{\partial}{\partial \theta} + \frac{i}{\sin\theta}
\frac{\partial}{\partial \phi} \right),
\end{eqnarray}
where $\delta_0 = \frac{\partial}{\partial \theta} +
\frac{i}{\sin\theta} \frac{\partial}{\partial \phi}$. The
associated dual reference tetrad is
\begin{eqnarray}
{\bf \stackrel\circ l} &=& - dU,
\nonumber\\
{\bf \stackrel\circ n} &=& - \frac12 \left( 1 + \frac{|\delta_0
\omega|^2}{(R + \omega)^2} \right) dU - dR,
\nonumber\\
{\bf \stackrel\circ m} &=& - \frac{{\rm e}^{-i\phi} \, \delta_0
\omega}{\sqrt2 (R + \omega)} dU + \frac1{\sqrt2} {\rm e}^{-i\phi}
(R + \omega)(d\theta + i \sin\theta d\phi).
\end{eqnarray}

From the above metric, a direct calculation gives the
non-vanishing NP reference coefficients as
\begin{eqnarray}
{\stackrel\circ\rho} &=& - \frac1{R + \omega},
\\
{\stackrel\circ\mu} &=& - \frac12 \left[ \frac1{R+\omega} +
\frac{|\delta_0 \omega|^2}{(R+\omega)^3} - \frac{\partial_\theta^2
\omega + \cot\theta \, \partial_\theta \omega +
\frac1{\sin^2\theta} \,
\partial_\phi^2 \omega}{(R+\omega)^2} \right],
\\
{\stackrel\circ\alpha} &=& - \frac{{\rm e}^{i\phi}}{2\sqrt2}
\left[ \frac{\cot\frac{\theta}2}{R+\omega} + \frac{2 \bar \delta_0
\omega}{(R+\omega)^2} \right],
\\
{\stackrel\circ{\bar\beta}} &=& \frac{{\rm e}^{i\phi}}{2\sqrt2}
\frac{\cot\frac{\theta}2}{R+\omega},
\\
{\stackrel\circ{\bar \tau}} &=& {\stackrel\circ\pi} = - \frac{{\rm
e}^{i\phi} \, \bar \delta_0 \omega}{\sqrt{2} (R + \omega)^2},
\\
{\stackrel\circ\gamma} &=& - \frac12 \left[ \frac{|\delta_0
\omega|^2}{(R+\omega)^3} + \frac{i \cot\frac{\theta}2 \,
\partial_\phi \omega}{(R+\omega)^2 \sin\theta} \right],
\\
{\stackrel\circ\lambda} &=& \frac{{\rm e}^{2i\phi}}{2(R+\omega)^2}
\left[ \partial_\theta^2 \omega - \frac{\partial_\phi^2
\omega}{\sin^2\theta} - \cot\theta \partial_\theta \omega +
\frac{2 i \cot\theta \partial_\phi \omega}{\sin\theta} - \frac{2
i}{\sin\theta} \partial_\theta \partial_\phi \omega -
\frac2{R+\omega} \left( \partial_\theta \omega - \frac{i
\partial_\phi \omega}{\sin\theta} \right)^2
\right]
\nonumber\\
&=& \frac{{\rm e}^{2i\phi}}{2(R+\omega)^2} \left[ \bar\delta_0^2
\omega - \cot\theta \, \bar\delta_0 \omega - \frac2{R+\omega} (
\bar\delta_0 \omega)^2 \right],
\\
{\stackrel\circ\nu} &=& - \frac{{\rm e}^{i\phi}}{\sqrt2
(R+\omega)^3} \left[ \frac{|\delta_0 \omega|^2 \, \bar\delta_0
\omega}{R+\omega} - \delta_0 \omega \left( \bar\delta_0
\partial_\theta \omega + \frac{i \cot\theta \partial_\phi
\omega}{\sin\theta} \right) + \frac{i \partial_\phi
\omega}{\sin\theta} \left( \partial_\theta^2 \omega + \cot\theta
\, \partial_\theta \omega + \frac1{\sin^2\theta} \,
\partial_\phi^2 \omega \right) \right].
\end{eqnarray}

The intrinsic geometry of the two sphere is preserved in the OST
embedding, therefore $\Delta \eta_{ab} \cong 0$. Moreover, on the
considered two sphere $S$, $R = 0$ by the definition of the
embedding and, from Eqs. (\ref{g2}) and (\ref{vol2}), we have
\begin{equation}
\omega(\theta, \phi) = r - \frac{|\sigma^0|^2}{2r} + O(r^{-2}).
\end{equation}
Therefore, this reference contribution differs from the result of
the holonomic embedding only by higher orders of $1/r$. Hence the
energy-momentum at null infinity in the OST embedding is the same
as in the holonomic embedding, i.e.
\begin{equation}
\lim_{\cal I^+}p_\nu = - \frac1{4 \pi} \int_S {\rm Re} \left(
\Psi^0_2 + \sigma^0 \dot {\bar \sigma}{}^0 + {\not\!\partial}_0^2
{\bar \sigma}^0 \right) f_\nu \, d\Omega^2.
\end{equation}
Again the result differs from the usual Bondi integrand by the
same extra term---which makes a vanishing contribution to the
energy when integrated over the 2-sphere.

\subsection{Brown-Lau-York embedding}
In the above two subsections, the two methods used both embedded
the surface $S$ into a standard light cone in Minkowski
space-time, i.e. the light cone $N$ is the light cone from one
point. Brown, Lau and York \cite{BLY97} gave another way to do the
embedding near null infinity. This method considers a more general
light cone.

Suppose we choose a Bondi coordinate system $(u, R, \theta, \phi)$
in Minkowski space-time. The asymptotic shear is
${\stackrel\circ\sigma}{}^0$. We can also do the formal Taylor
extension near null infinity as in Eqs.
(\ref{asyrho}--\ref{asytau}). The only difference is that we have
$\stackrel\circ\Psi_i = 0, \, n=0,1,2,3,4$. From the NP equations
we have
\begin{equation}
\Psi^{(0)}_3 = - {\not\!\partial}_0 {\partial_u
{\stackrel\circ{\bar\sigma}}}{}^0, \qquad \Psi^{(0)}_4 = -
\partial_u^2 {\stackrel\circ{\bar\sigma}}{}^0.
\end{equation}
 Because the spin-weight of $\partial_u
{\stackrel\circ{\bar\sigma}}{}^0$ is non-zero, the above results
tell us that $\partial_u {\stackrel\circ{\bar\sigma}}{}^0 = 0$
\cite{St90}. Insert this results into Eqs.
(\ref{asyrho}--\ref{asytau}), the N-P quantities in Minkowski
space-time are then
\begin{eqnarray}
\stackrel\circ\rho &=& - \frac1{R} -
\frac{|\stackrel\circ\sigma{}^0|^2}{R^3} + O(R^{-5}),
\label{BLYrho} \nonumber\\
\stackrel\circ\sigma &=& \frac{\stackrel\circ\sigma{}^0}{R^2} +
\frac{|\stackrel\circ\sigma{}^0|^2 \,
\stackrel\circ\sigma{}^0}{R^4} + O(R^{-5}),
\nonumber\\
\stackrel\circ\alpha &=& \frac{\alpha^0}{R} +
\frac{\stackrel\circ{\bar\sigma}{}^0 \, \bar \alpha^0 +
{\not\!\partial}_0 \stackrel\circ{\bar\sigma}{}^0}{R^2} +
O(R^{-3}),
\nonumber\\
\stackrel\circ\beta &=& - \frac{\bar \alpha^0}{R} - \frac{\alpha^0
\stackrel\circ\sigma{}^0}{R^2} + O(R^{-3}),
\nonumber\\
\stackrel\circ\tau &=& \stackrel\circ{\bar\pi} = \frac{\bar
{\not\!\partial}_0 \stackrel\circ\sigma{}^0}{R^2} + O(R^{-3}),
\nonumber\\
\stackrel\circ\gamma &=& \frac{\stackrel\circ\gamma_1}{R^2} +
O(R^{-3})
= \frac{-\alpha^0 \bar {\not\!\partial}_0 \stackrel\circ\sigma{}^0
+ \bar \alpha^0 {\not\!\partial}_0 \stackrel\circ{\bar\sigma}{}^0}
{R^2} + O(R^{-3}),
\nonumber\\
\stackrel\circ\mu &=& - \frac{1}{2R} - \frac{{\not\!\partial}_0^2
\stackrel\circ{\bar\sigma}{}^0}{R^2} + O(R^{-3}),
\nonumber\\
\stackrel\circ\lambda &=& \frac{\frac12
\stackrel\circ{\bar\sigma}{}^0 - \bar {\not\!\partial}_0
{\not\!\partial}_0 \stackrel\circ{\bar\sigma}{}^0}{R^2} +
O(R^{-3}),
\nonumber\\
\stackrel\circ\nu &=& O(R^{-3}). \label{BLYnu}
\end{eqnarray}
The tetrad part is
\begin{eqnarray}
\stackrel\circ U &=& - \frac12 - \frac{\stackrel\circ\gamma_1 +
\stackrel\circ{\bar\gamma}_1}{R} + O(R^{-2}),
\nonumber\\
\stackrel\circ X &=& - \frac{\bar {\not\!\partial}_0
\stackrel\circ\sigma{}^0 \xi^0}{R^2} + O(R^{-3}),
\nonumber\\
\stackrel\circ\xi &=& \frac{\xi^0}{R} +
\frac{|\stackrel\circ\sigma{}^0|^2 \xi^0}{R^3} + O(R^{-4}),
\nonumber\\
\stackrel\circ\eta &=& - \frac{\stackrel\circ{\bar\sigma}{}^0
\xi^0}{R^2} + O(R^{-4}).
\end{eqnarray}

The intrinsic geometry of the 2-sphere is preserved in the
embedding, i.e.
\begin{equation}
{\cal R} = {\stackrel\circ{\cal R}},
\end{equation}
where the two dimensional Ricci scalar ${\cal R}$  is given by
${\cal R} = - 2 \rho \mu - 2 \bar \rho \bar \mu + 2 \sigma \lambda
+ 2 \bar \sigma \bar \lambda + 2 \Psi_2 + 2 \bar \Psi_2$.
Consequently, this leads to
\begin{equation}
- \mu \rho + \lambda \sigma + \Psi_2 + \bar\Psi_2 + \bar\lambda
\bar\sigma - \bar\mu \bar\rho = - \stackrel\circ\mu
\stackrel\circ\rho - \stackrel\circ{\bar\mu}
\stackrel\circ{\bar\rho} + \stackrel\circ\lambda
\stackrel\circ\sigma + \stackrel\circ{\bar\lambda}
\stackrel\circ{\bar\sigma}.
\end{equation}
Using the Taylor expansion in Eqs. (\ref{asyrho}--\ref{asytau})
and Eqs. (\ref{BLYnu}), we find that the relation between the
parameter $r$ and $R$ is
\begin{equation}
R = r + k + O(r^{-1}), \quad {\rm with} \quad k =
{\not\!\partial}_0^2 {\stackrel\circ{\bar\sigma}}{}^0 +
{\bar{\not\!\partial}}_0^2 {\stackrel\circ\sigma}{}^0 -
{\not\!\partial}_0^2 {\bar\sigma}^0 - {\bar{\not\!\partial}}_0^2
\sigma^0.
\end{equation}
We choose ${\stackrel\circ\sigma}{}^0|_{S_0}=\sigma^0|_S $, where
$S$ is the section on ${\cal I}^+$ and $S_0$ is its image under
the embedding. We have
\begin{equation}
R = r + O(r^{-1}) = r [1 + O(r^{-2})].
\end{equation}

As for the OST embedding, the two sphere geometry is preserved,
therefore $\Delta \eta_{ab} \cong 0$. Finally the gravitational
energy-momentum in the BLY embedding is
\begin{equation}
\lim_{\cal I^+} p_\nu = - \frac1{4 \pi} \int_S {\rm Re} \left(
\Psi^0_2 + \sigma^0 \dot {\bar \sigma}{}^0 \right) f_\nu \,
d\Omega^2.
\end{equation}
It is worth noting that the BLY embedding is a little neater, in
that it directly gives the standard Bondi energy-momentum
integrand, whereas there is an additional term in the other two
embeddings (which vanishes upon integration). That term is
associated with the embedding methods in which the section $S$ is
not embedded into a standard light cone, generated by the null
geodesics starting from a single point. This difference again
shows us that keeping the inner geometry unchanged under the
embedding is not enough to ensure physically reasonable
quasi-local quantities; generally, as is especially clear from
\cite{OST04}, we need more restrictions.

\section{The energy flux at null infinity via the direct method}
In this section, we directly calculate the energy flux through a
two sphere. For simplicity we choose the time-like translation to
be $N = n + \frac12 l$. Asymptotically, this agrees with the
natural choice, the time translation of the BMS group at null
infinity $\partial_u$.  The difference between the two vectors is
asymptotically negligible.

We consider the energy $E = H(N,S)$. Suppose $\Sigma_0$ is the
space-like region that we want to consider, $\partial \Sigma_0 =
S$, and $\Sigma_{\Delta t}$ is the time evolution of $\Sigma_0$.
The energy within the region during the time interval changes by
the amount $E(\Sigma_{\Delta t}) - E(\Sigma_0)$, hence a natural
direct definition of the rate of energy change is
\begin{equation}
\dot E := \lim_{\Delta t \to 0} \frac{E(\Sigma_{\Delta t}) -
E(\Sigma_0)}{\Delta t}.
\end{equation}
Looking to the value of the Hamiltonian, taking into account the
vanishing of the initial value constraints, from (\ref{Energy}) we
straightforwardly get the quasilocal energy flux relation
\begin{eqnarray}
\dot E = \dot H(N,\Sigma) &:=& \frac1{16\pi} \oint_S \Lie_N \left(
\Delta \omega^{ab} \wedge i_N \eta_{ab} -
{\stackrel\circ\nabla}{}^a {\stackrel\circ N}{}^b \Delta \eta_{ab}
\right)
\nonumber \\
&=& \frac1{16\pi} \oint_S \left[ \Lie_N \left( \omega^{ab} \wedge
i_N \eta_{ab} \right) - \Lie_N \left(
{\stackrel\circ\omega}{}^{ab} \wedge i_N \eta_{ab} \right) -
\Lie_N \left( {\stackrel\circ\nabla}{}^a {\stackrel\circ N}{}^b
\Delta \eta_{ab} \right) \right].
\end{eqnarray}
The right hand side defines an energy flux expression $F$ which
includes two parts that will be considered separately: the purely
physical part and the other part, which includes the reference,
i.e.\ $F = F_{\rm phy} + F_{\rm ref}$. The reason for making such
a separation is that the part including the reference, as we have
already seen, depends on the embedding of the reference
configuration into physical space-time.

We first evaluate the physical part of the flux,
\begin{equation}
\Lie_N ( \omega^{ab} \wedge i_N \eta_{ab}) = \frac12 \Lie_l (
\omega^{ab} \wedge i_N \eta_{ab}) + \Lie_n (\omega^{ab} \wedge i_N
\eta_{ab} ).
\end{equation}
We see that it is necessary to know the Lie derivative of all two
form elements. However, we are only interested in the final
results that can contribute to the integral: the terms
proportional to ${\bf m} \wedge {\bf \bar m}$. After a
straightforward verification the contributing terms are
\begin{equation}
\Lie_l ( {\bf m} \wedge {\bf \bar m} ) \cong - ( \rho + \bar \rho)
\, {\bf m} \wedge {\bf \bar m}, \qquad \Lie_n ( {\bf m} \wedge
{\bf \bar m} ) \cong ( \mu + \bar \mu) \, {\bf m} \wedge {\bf \bar
m}, \qquad \Lie_n ( {\bf n} \wedge {\bf m} ) \cong - \bar \nu \,
{\bf m} \wedge {\bf \bar m}.
\end{equation}

Therefore, following the result of $\omega^{ab} \wedge i_N
\eta_{ab}$ in Eq.~(\ref{Freud}), the first term is
\begin{eqnarray}
\Lie_l ( \omega^{ab} \wedge i_N \eta_{ab}) &\cong& i D \left[
(\rho + \bar \rho) + 2(\mu + \bar \mu) \right] \, {\bf m} \wedge
{\bf \bar m} + i \left[ (\rho + \bar \rho) + 2(\mu + \bar \mu)
\right] \, \Lie_l ( {\bf m} \wedge {\bf \bar m} )
\nonumber\\
&\cong& i \left\{ D \left[ (\rho + \bar \rho) + 2(\mu + \bar \mu)
\right] - ( \rho + \bar \rho) \left[ (\rho + \bar \rho) + 2(\mu +
\bar \mu) \right] \right\} \, {\bf m} \wedge {\bf \bar m}
\nonumber\\
&=& i \left[ 2 \delta \pi + 2 \bar \delta \bar \pi + {\cal R} - 2
\rho^2 + 2 |\sigma|^2 + 4 |\pi|^2 - 2 (\bar \alpha - \beta) \pi -
2 (\alpha - \bar\beta) \bar\pi \right] \, {\bf m} \wedge {\bf \bar
m},
\end{eqnarray}
where ${\cal R}$ is the two dimensional Ricci scalar on the
enclosed surface. Similarly, the second term is
\begin{eqnarray}
\Lie_n ( \omega^{ab} \wedge i_N \eta_{ab}) &\cong& i D' \left[
(\rho + \bar \rho) + 2(\mu + \bar \mu) \right] \, {\bf m} \wedge
{\bf \bar m} + i \left[ (\rho + \bar \rho) + 2(\mu + \bar \mu)
\right] \, \Lie_n ( {\bf m} \wedge {\bf \bar m} )
\nonumber\\
&& - i 2 (\alpha - \bar \beta - \pi) \Lie_n ( {\bf n} \wedge {\bf
m} ) + i 2 (\bar \alpha - \beta - \bar \pi) \Lie_n ( {\bf n}
\wedge {\bf \bar m} )
\nonumber\\
&\cong& i \left\{ D' \left[ (\rho + \bar \rho) + 2(\mu + \bar \mu)
\right] + ( \mu + \bar \mu) \left[ (\rho + \bar \rho) + 2(\mu +
\bar \mu) \right] \right.
\nonumber\\
&& \left. + 2 (\alpha - \bar \beta - \pi) \bar \nu - 2 (\bar
\alpha - \beta - \bar \pi) \nu \right\} \, {\bf m} \wedge {\bf
\bar m}
\nonumber\\
&=& i \Bigl[\delta \bar \tau + \bar \delta \tau + 2 \delta \nu + 2
\bar \delta \bar \nu - \frac12 {\cal R} + 4 \mu^2 - 4 |\lambda|^2
+ 2 (\gamma + \bar \gamma)(\rho - 2 \mu) + 2 |\pi|^2 + (\bar
\alpha - \beta) \pi
\nonumber\\
&& + (\alpha - \bar \beta) \bar \pi + 2 \nu (\bar \alpha + 3
\beta) + 2 \bar \nu (\alpha + 3 \bar \beta) + 2 (\alpha - \bar
\beta - \pi) \bar \nu - 2 (\bar \alpha - \beta - \bar \pi) \nu
\Bigr] \, {\bf m} \wedge {\bf \bar m},
\end{eqnarray}
Finally, the purely physical contribution to the energy flux is
\begin{eqnarray}
F_{\rm phy} &=& \frac1{16 \pi} \int_S \Lie_N ( \omega^{ab} \wedge
i_N \eta_{ab} )
\nonumber\\
&=& \frac{i}{8 \pi} \int_S \Bigl[ \delta \pi + \bar \delta \bar
\pi + \delta \nu + \bar \delta \bar \nu - \frac12 \rho^2 + \frac12
|\sigma|^2 + 2 \mu^2 + 2 |\pi|^2 - 2 |\lambda|^2
\nonumber\\
&& + (\gamma + \bar \gamma)(\rho - 2 \mu) + \nu (\bar \alpha + 3
\beta) + \bar \nu (\alpha + 3 \bar \beta) + (\alpha - \bar \beta -
\pi) \bar \nu - (\bar \alpha - \beta - \bar \pi) \nu \Bigr] \,
{\bf m} \wedge {\bf \bar m}.
\end{eqnarray}
From the asymptotic expansion of all the NP coefficients
(\ref{asyrho}-\ref{asytau}, \ref{asygamma}, \ref{asymu},
\ref{asylambda}, \ref{asynu}) we find that all except one of the
terms fall off as $O(1/r^3)$ or faster, only the $-2|\lambda|^2$
term contributes asymptotically. Hence the null infinity limit of
the purely physical flux is
\begin{eqnarray}
\lim_{\cal I^+} F_{\rm phy} &=& \lim_{\cal I^+} \frac1{8\pi}
\int_S \left[ - 2 \frac{|{\dot {\bar \sigma}}{}^0|^2}{r^2} +
O(r^{-3}) \right] \, i \ {\bf m} \wedge {\bf \bar m}
\nonumber\\
&=& - \frac1{4\pi} \int_S|{\dot {\bar \sigma}}{}^0|^2 d\Omega^2.
\end{eqnarray}

\subsection{Holonomic embedding}
A straightforward calculation gives
\begin{eqnarray}
\Lie_N \left( {\stackrel\circ\omega}{}^{ab} \wedge i_N \eta_{ab}
\right) &=& \Lie_N \left[ ( \stackrel\circ\rho +
\stackrel\circ{\bar\rho} + 2 \stackrel\circ\mu + 2
\stackrel\circ{\bar\mu} ) \, i \ {\bf m} \wedge {\bf \bar m} +
(\stackrel\circ{\bar\alpha} - \stackrel\circ\beta) \, i \ {\bf n}
\wedge {\bf \bar m} - (\stackrel\circ\alpha -
\stackrel\circ{\bar\beta}) \, i \ {\bf n} \wedge {\bf m} + O(r^{-3})
\right]
\nonumber\\
&\cong& \left[ (\stackrel\circ\rho + \stackrel\circ{\bar\rho} + 2
\stackrel\circ\mu + 2 \stackrel\circ{\bar\mu} )(- \frac12 \rho -
\frac12 \bar\rho + \mu + \bar\mu) + O(r^{-3}) \right] \, i \ {\bf m}
\wedge {\bf \bar m},
\nonumber\\
\Lie_N \left( {\stackrel\circ\nabla}{}^a {\stackrel\circ N}{}^b
\Delta \eta_{ab} \right) &=& \Lie_N \left[ - \frac12
(\stackrel\circ\gamma + \stackrel\circ{\bar\gamma}) \, i \ ( {\bf m}
\wedge {\bf \bar m} - \stackrel\circ{\bf m} \wedge
\stackrel\circ{\bf \bar m} ) + \frac12 \, \stackrel\circ{\bar\nu} \,
i \ ( {\bf l} \wedge {\bf \bar m} - \stackrel\circ{\bf l} \wedge
\stackrel\circ{\bf \bar m} ) - \frac12 \stackrel\circ\nu \, i \ (
{\bf l} \wedge {\bf m} - \stackrel\circ{\bf l} \wedge
\stackrel\circ{\bf m}) \right]
\nonumber\\
&\cong& 0.
\end{eqnarray}
Therefore, the reference part in the holonomic embedding is
\begin{equation}
F_{\rm ref} = - \frac1{16\pi} \int_S \left[ ( \stackrel\circ\rho +
\stackrel\circ{\bar\rho} + 2 \stackrel\circ\mu + 2
\stackrel\circ{\bar\mu} ) ( - \frac12 \rho - \frac12 \bar\rho + \mu
+ \bar\mu ) + O(r^{-3}) \right] \, i \ {\bf m} \wedge {\bf \bar m} =
\frac1{16\pi} \int_S O(r^{-3}) \, i \ {\bf m} \wedge {\bf \bar m}.
\end{equation}
Finally, we have
\begin{equation}
\lim_{\cal I^+}F = - \frac1{4\pi} \int_S |{\dot\sigma}^0|^2 \,
d\Omega^2. \label{Eflux}
\end{equation}
This is just the standard expression of the Bondi energy flux at
${\cal I}^+$ \cite{PR84,St90,NT80,Haw68}.

\subsection{\'O Murchadha-Szabados-Tod embedding}
The reference part in the OST embedding is
\begin{eqnarray}
\Lie_N \left( {\stackrel\circ\omega}{}^{ab} \wedge i_N \eta_{ab}
\right) &\cong& \left[ (- \stackrel\circ{\bar\pi} +
\stackrel\circ{\bar\alpha} - \stackrel\circ\beta) \nu + (-
\stackrel\circ\pi + \stackrel\circ\alpha -
\stackrel\circ{\bar\beta}) \bar\nu + (\stackrel\circ\rho +
\stackrel\circ{\bar\rho} + \stackrel\circ\mu +
\stackrel\circ{\bar\mu})(- \frac{\rho + \bar\rho}2 + \mu + \bar\mu)
\right] \, i \ {\bf m} \wedge {\bf \bar m},
\nonumber\\
\Lie_N \left( {\stackrel\circ\nabla}{}^a {\stackrel\circ N}{}^b
\Delta \eta_{ab} \right) &\cong& {\stackrel\circ\nabla}{}^a
{\stackrel\circ N}{}^b \Lie_N \left( \Delta \eta_{ab} \right) \cong
- \left[ ( \stackrel\circ\gamma + \stackrel\circ{\bar\gamma} )
\Lie_N \Delta ( {\bf m} \wedge {\bf \bar m} ) + \stackrel\circ\nu
\Lie_N \Delta ( {\bf l} \wedge {\bf m} ) - \stackrel\circ{\bar\nu}
\Lie_N \Delta ( {\bf l} \wedge {\bf \bar m} ) \right].
\end{eqnarray}

The image of $S$ is the two sphere in Minkowski space-time such
that $U = {\rm const}.$ and $R=0$. From the above calculation, we
get the same result as before:
\begin{equation}
F_{\rm ref} = \int_S O(r^{-3}) \, i \ {\bf m} \wedge {\bf \bar m}.
\end{equation}
Consequently, for the energy flux we get the same result as Eq.
(\ref{Eflux}).

\subsection{Brown-Lau-York embedding}
Following up on the result in the case of the energy-momentum
calculation, the reference part of the energy flux is
\begin{equation}
F_{\rm ref} = \int_S O(r^{-3}) i {\bf m} \wedge {\bf \bar m}.
\end{equation}
Hence for the total energy flux we again we get the same result as
Eq. (\ref{Eflux}).

\section{The energy flux at null infinity via an identity}
In this section, we calculate the energy flux through a two sphere
using an interesting formal Hamiltonian identity \cite{Nes84}
derived in detail in \cite{CNT05}.  The identity is simply the
analogue of the classical mechanics identity $\dot H \equiv 0$,
which follows from $\delta H=\dot q^k \delta p_k - \dot p_k \delta
q_k$ by simply by replacing $\delta\to d/dt$. We can obtain it
directly in essentially the same way from (\ref{deltaH}) simply by
substituting the time derivative operator $ \Lie_N$ for $\delta$;
then the field equation terms cancel identically (just as they did
in the classical mechanics case) leaving
\begin{equation}
\dot E := \dot H(N,\Sigma) \equiv - \oint_S i_N (\Delta\omega^{ab}
\wedge \Lie_N \eta_{ab}) \equiv \oint_S (- i_N \Delta \omega^{ab} \,
\Lie_N \eta_{ab} + \Delta\omega^{ab} \wedge \Lie_N i_N \eta_{ab})
\label{ef2}.
\end{equation}
This is a general quasi-local formula for energy flux, applicable
to the boundary of any region. In \cite{CNT05} it was tested in
the null infinity limit using the Bondi-Sachs metric.

Here we wish to transcribe this expression into the NP spin
coefficient form and confirm that it gives the desired asymptotic
results using that well developed technique. For our calculation
here we take $N= \partial_u = {\stackrel\circ n} + \frac12
{\stackrel\circ l} = n + \frac12 l + O(1/r)$, the reference geometry
Killing field.

Let us consider Eq. (\ref{ef2}) term by term. The first term is
\begin{eqnarray}
i_N \Delta \omega^{ab} \, \Lie_N \eta_{ab} &=& \Delta \omega_n^{ab}
\Lie_n \eta_{ab} + \frac12 \Delta \omega_n^{ab} \Lie_l \eta_{ab} +
\frac12 \Delta \omega_l^{ab} \Lie_n \eta_{ab} + \frac14 \Delta
\omega_l^{ab} \Lie_l \eta_{ab}
\nonumber\\
&=& 2 \Delta (\gamma + \bar\gamma) \Lie_n (i \ {\bf m} \wedge {\bf
\bar m}) - 2 \Delta \nu \Lie_n (i \ {\bf m} \wedge {\bf l}) - 2
\Delta \bar\nu \Lie_n (i \ {\bf l} \wedge {\bf \bar m})
\nonumber\\
&& + 2 \Delta \bar\tau \Lie_n (i \ {\bf n} \wedge {\bf m}) + 2
\Delta \tau \Lie_n (i \ {\bf \bar m} \wedge {\bf n}) + 2 \Delta
(\gamma - \bar\gamma) \Lie_n (i \ {\bf n} \wedge {\bf l})
\nonumber\\
&& + \Delta (\gamma + \bar\gamma) \Lie_l (i \ {\bf m} \wedge {\bf
\bar m}) - \Delta \nu \Lie_l (i \ {\bf m} \wedge {\bf l}) - \Delta
\bar\nu \Lie_l (i \ {\bf l} \wedge {\bf \bar m})
\nonumber\\
&& + \Delta \bar\tau \Lie_l (i \ {\bf n} \wedge {\bf m}) + \Delta
\tau \Lie_l (i \ {\bf \bar m} \wedge {\bf n}) + \Delta (\gamma -
\bar\gamma) \Lie_l (i \ {\bf n} \wedge {\bf l})
\nonumber\\
&& - \Delta (\varepsilon + \bar\varepsilon) \Lie_n (i \ {\bf m}
\wedge {\bf \bar m}) - \Delta \pi \Lie_n (i \ {\bf m} \wedge {\bf
l}) - \Delta \bar\pi \Lie_n (i \ {\bf l} \wedge {\bf \bar m})
\nonumber\\
&& - \Delta \bar\kappa \Lie_n (i \ {\bf n} \wedge {\bf m}) - \Delta
\kappa \Lie_n (i \ {\bf \bar m} \wedge {\bf n}) + \Delta
(\varepsilon - \bar\varepsilon) \Lie_n (i \ {\bf n} \wedge {\bf l})
\nonumber\\
&& + \frac12 \bigl[ - \Delta (\varepsilon + \bar\varepsilon) \Lie_l
(i \ {\bf m} \wedge {\bf \bar m}) - \Delta \pi \Lie_l (i \ {\bf m}
\wedge {\bf l}) - \Delta \bar\pi \Lie_l (i \ {\bf l} \wedge {\bf
\bar m})
\nonumber\\
&& \qquad - \Delta \bar\kappa \Lie_l (i \ {\bf n} \wedge {\bf m}) -
\Delta \kappa \Lie_l (i \ {\bf \bar m} \wedge {\bf n}) + \Delta
(\varepsilon - \bar\varepsilon) \Lie_l (i \ {\bf n} \wedge {\bf l})
\bigr].
\end{eqnarray}
Based on the asymptotic estimations given in the previous sections
and the three considered embedding methods, all of these terms are
of higher order than $O(1/r)$, hence they make no contribution to
the flux asymptotically. Basically this comes about because (i) in
our gauge $\epsilon=\kappa=0$, (ii) the Lie derivative introduces
a factor of $1/r$, (iii) the spin coefficients
$\sigma,\pi,\tau,\gamma,\nu$ are  $O(1/r^2)$, (iii) for
$\rho,\alpha,\beta,\mu$ the  $\Delta$ operation removes their
$O(1/r)$ part.

Similarly, for the second term we find
\begin{eqnarray}
\Delta \omega^{ab} \wedge \Lie_N i_N \eta_{ab} &\cong& (-\Delta
\omega_m^{ab} i_{\bar m} \Lie_N i_N \eta_{ab} + \Delta \omega_{\bar
m}^{ab} i_m \Lie_N i_N \eta_{ab}) {\bf m} \wedge {\bf \bar m}
\nonumber\\
&=& 2 \bigl[ \Delta (\bar\alpha + \beta) i_{\bar m} \Lie_N i_N
\eta_{01} - (\Delta \mu) i_{\bar m} \Lie_N i_N \eta_{02} - (\Delta
\bar\lambda) i_{\bar m} \Lie_N i_N \eta_{03}
\nonumber\\
&& + (\Delta \bar\rho) i_{\bar m} \Lie_N i_N \eta_{12} + (\Delta
\sigma) i_{\bar m} \Lie_N i_N \eta_{13} + \Delta (\bar\alpha -
\beta) i_{\bar m} \Lie_N i_N \eta_{23}
\nonumber\\
&& - \Delta (\alpha + \bar\beta) i_m \Lie_N i_N \eta_{01} + (\Delta
\lambda) i_m \Lie_N i_N \eta_{02} + (\Delta \bar\mu) i_m \Lie_N i_N
\eta_{03}
\nonumber\\
&& - (\Delta \bar\sigma) i_m \Lie_N i_N \eta_{12} - (\Delta \rho)
i_m \Lie_N i_N \eta_{13} + \Delta (\alpha - \bar\beta) i_m \Lie_N
i_N \eta_{23} \bigr] {\bf m} \wedge {\bf \bar m}
\nonumber\\
&=& - 4 |\lambda|^2 \, i \ {\bf m} \wedge {\bf \bar m} +
O(\frac1{r}).
\end{eqnarray}
Again we used the asymptotic estimations and the three considered
embedding methods. In this case we get a non-vanishing asymptotic
contribution from $\lambda = O(1/r)$ (\ref{asylambda}). Submitting
these results into Eq. (\ref{ef2}), we find that the null infinity
limit of the energy flux calculated from that Hamiltonian identity
relation is
\begin{equation}
\lim_{\cal I^+} F = - \frac1{4\pi} \int_S |{\dot\sigma}^0|^2 \,
d\Omega^2,
\end{equation}
which is, just as was found from the direct calculation
(\ref{Eflux}), the standard flux loss due to the Bondi news.

\section{Discussion}

We have tested certain expressions for the quasilocal
energy-momentum and energy flux of gravitating systems.  The
expressions were obtained from the covariant Hamiltonian
formalism. In this formalism the quasilocal quantities are
determined by the value of the boundary term in the Hamiltonian.
The variation of the Hamiltonian associates the choice of boundary
term with specific boundary conditions. Thus the definition of the
quasilocal energy-momentum of a gravitating system is linked to
the choice of boundary conditions.  The boundary term that
corresponds to holding certain projected components of the
orthonormal frame fixed seems to be the best choice for most
purposes. We have considered only that choice here (the values for
certain other choices are given in \cite{CNT05}).

Energy flux can be computed in more than one way.  On the one hand
it can be obtained directly from the change in the energy
expression.  On the other hand one can use an interesting identity
associated with the specific role of the Hamiltonian and its
variation.   Here we have evaluated the energy flux by both
techniques.

In strong field regions we do not have any sharp test as to what
values we should find for energy-momentum and energy flux.
Proposed expressions necessarily are first tested in the weak
field linearized theory limits.  Getting good values at spatial
infinity is not the strongest test.  The Bondi limit at future
null infinity is more delicate.  Here we tested our selected
expression for energy-momentum and its associate energy flux in
this limit.

Technically we used a well known and well developed technique: the
Newman-Penrose spin coefficients.  We selected a suitable gauge
and found that we needed certain quantities expanded in more
detail than is usual \cite{Sha86}. In the quasilocal expressions
it is necessary to select reference values which determine the
``vacuum'' or ``ground state''. The natural choice is, of course,
Minkowski space, but it is not so obvious how to embed the
Minkowski space into the asymptotic part of the dynamic space.  We
considered three types of embeddings which have been used:
holonomic, one due to \'O~Murchadha, Szabados and Tod
\cite{OST04}, and one due to Brown, York and Lau \cite{BLY97}. We
found some interesting technical differences between the
embeddings but in the end they all gave the same answer: namely
the expected Bondi energy and the Bondi energy flux determined by
the Bondi news.

In the detailed calculation we noted that, at least in the
selected gauge, the quasilocal energy was asymptotically
determined by the deviation of the spin coefficient $\mu$ from its
asymptotic Minkowski value and that the energy flux was determined
by the spin coefficient $\lambda$ (in the notation of \cite{GHP73}
these coefficients are $-\rho'$ and $-\sigma'$, respectively).

We have shown that the values of these expressions can be
practically calculated in terms of the NP spin coefficient
technique; the expressions were found to have the desired
asymptotic values. Thus they satisfy an important criterion for
quasi-local energy and energy flux expressions.

\section*{Acknowledgements}
This work was supported by grants from the National Science Council
of the Republic of China,  X.~Wu was supported by grant number NSC
92-2816-M-008-0004-6, C.-M.~Chen by grant number NSC
93-2112-M-008-021, and J.~M.~Nester by grant number NSC
93-2112-M-008-001.



\end{document}